\documentclass[pdftex,twocolumn,epjc3]{svjour3}          % twocolumn

\RequirePackage[T1]{fontenc}

\smartqed  % flush right qed marks, e.g. at end of proof

\RequirePackage{graphicx}
\RequirePackage{mathptmx}      % use Times fonts if available on your TeX system
\usepackage{amssymb}
\usepackage{amsmath}
\usepackage{changes}
\usepackage{float}
\usepackage{epstopdf}
\usepackage{longtable}

\def\mnras{MNRAS}
\def\apj{ApJ}
\def\prd{PRD}
\def\jcap{JCAP}

\def\aap{A\&A}

\RequirePackage{flushend}
\RequirePackage[numbers,sort&compress]{natbib}
\RequirePackage[colorlinks,citecolor=blue,urlcolor=blue,linkcolor=blue]{hyperref}
\journalname{Eur. Phys. J. C}

\begin{document}

\title{Gravitational waves from pulsars in the context of magnetic ellipticity}


\author{Jos\'e C. N. de Araujo\thanksref{e1}
        \and
        Jaziel G. Coelho\thanksref{e2}
        \and Cesar A. Costa\thanksref{e3}
}

%\thankstext[$\star$]{t1}{Thanks to the title}
\thankstext{e1}{e-mail:jcarlos.dearaujo@inpe.br}
\thankstext{e2}{e-mail:jaziel.coelho@inpe.br}
\thankstext{e3}{e-mail:cesar.costa@inpe.br}
\institute{Divis\~ao de Astrof\'isica, Instituto Nacional de Pesquisas Espaciais, Avenida dos Astronautas 1758, S\~ao Jos\'e dos Campos, 12227--010 SP, Brazil\label{addr1}}

\date{Received: date / Accepted: date}

\maketitle

\begin{abstract}
In one of our previous articles we have considered the role of a time dependent magnetic ellipticity on pulsars' braking indices and on the putative gravitational waves these objects can emit. Since only nine of more than 2000 known pulsars have accurately measured braking indices, it is of interest to extend this study to all known pulsars, in particular to what concerns the gravitational waves generation. To do so, as shown in our previous article, we need to know some pulsars' observable quantities such as: periods and their time derivatives, and estimated distances to the Earth. Moreover, we also need to know the pulsars' masses and radii, for which we are adopting current fiducial values. Our results show that the gravitational wave amplitude is at best $h \sim 10^{-28}$. This leads to a pessimistic prospect for the detection of gravitational waves generated by these pulsars, even for Advanced LIGO and Advanced Virgo, and the planned Einstein Telescope, whether the ellipticity has magnetic origin. 
\end{abstract}

\section{Introduction}
\label{int}
It is well known that, besides compact binaries, rapidly rotating neutron stars are promising sources of gravitational waves (GWs) which could be detected in a near future by Advanced LIGO (aLIGO) and Advanced Virgo (AdV), and also by the planned Einstein Telescope (ET). These sources generate continuous GWs whether they are not perfectly symmetric around their rotation axis, i.e. if they present some equatorial ellipticity. 

It is worth stressing that the equatorial ellipticity is an extremely relevant parameter since the GW amplitude is directly proportional to it. Therefore, whether the ellipticity be extremely small, i.e., $\epsilon\ll 10^{-5}$, the GW amplitude will be also extremely small, implying that the detection of such continuous GWs generated by pulsars may be unattainable~\citep[see][]{2016arXiv160305975D,2016EPJC} with the current technology. For a matter of comparison, some authors argue that an acceptable upper limit for the ellipticity would be around $\epsilon\sim 10^{-6}$ \citep[see, e.g.,][]{2008PhLB..668....1K}. An important mechanism for producing asymmetries is the development of non-axisymmetric instabilities in rapidly rotating neutron stars driven by the gravitational emission reaction or by nuclear matter viscosity~\citep[see, e.g.,][and references therein]{1996ApJ...460..379B}. 

We explore, in the present paper, some consequences of an ellipticity generated by the magnetic dipole of the pulsars themselves. It is well known that, for strong magnetic fields ($\sim10^{12}-10^{15}$~G), the equilibrium configuration of a neutron star can be distorted due to the magnetic pressure. Therefore, both rotation and magnetic field combined can produce a flattened equilibrium star. However, star rotation and strong magnetic field may not be sufficient for GW emission, other effects must be associated, such as pulsar precession~\citep[see, e.g.,][]{1979PhRvD..20..351Z}.

The main goal of the present paper is to extend our previous study where the role of ellipticity of magnetic origin ($\epsilon_{B}$) was considered \citep{2016ApJ}. Since in that paper we were also interested in braking indices, which are until now accurately measured for only nine pulsars, we had restricted that study exclusively for those very pulsars.

However, $\epsilon_{B}$ does not depend on the pulsar braking index, maybe it is the other way around. In fact, $\epsilon_{B}$ is mostly associated to both the pulsar period ($P$) and its time derivative ($\dot{P}$), for a given value of mass, radius and moment of inertia. Therefore, it is straightforward to extend the $\epsilon_{B}$ calculation for all pulsars with known $P$ and $\dot{P}$. Consequently, with $\epsilon_{B}$ in hands, we can calculate the GW amplitudes for all pulsars with known $P$, $\dot{P}$, and their distances to the Earth.

To do so this paper is organized as follows. Section \ref{sec:2} is devoted to a brief procedure description which is conducted by a basic set of equations. In Section \ref{sec:3} we present how the calculations are done and discuss the obtained results. And, finally, in Section \ref{sec:4} we summarize the main conclusions and remarks about them. 

\section{Basic equations}
\label{sec:2}
In \citet{2016ApJ} we consider in detail how to relate $\epsilon_{B}$ to $P$ and $\dot{P}$. In addition, the basic equations used for calculating the amplitude of the putative GWs generated by pulsars are also presented. All those equations are used to calculate the relevant quantities of this present paper. Therefore, here we are only providing the main steps for deriving these relevant equations.

Recall that the equatorial ellipticity is given by
~\citep[see e.g.,][]{1983bhwd.book.....S}
\begin{equation}
\epsilon=\frac{I_{xx}-I_{yy}}{I_{zz}},
\end{equation}
where $I_{xx}$, $I_{yy}$, $I_{zz}$ are the moment of inertia with respect to the rotation axis, $z$, and along directions perpendicular to it.

Regarding the ellipticity of magnetic origin, it was shown by different authors \citep{1996A&A...312..675B,2000A&A...356..234K,2006A&A...447....1R} that it is given by
\begin{equation}
\epsilon_B = \kappa\frac{B_0^2 R^4}{G M^2}\sin^2\phi, \label{eq:epsilonB}
\end{equation}
where $B_0$ is the dipole magnetic field, $R$ and $M$ are the radius and the mass of the star respectively, $\phi$ is the angle between the rotation and magnetic dipole axes, whereas $\kappa$ is the distortion parameter, which depends on both the star equation of state (EoS) and the magnetic field configuration~\citep[see e.g.][]{2006A&A...447....1R}. 

Regarding the GW amplitude, one finds in the literature the following equation
\begin{equation}
h^{SD} = \left( \frac{5}{2}\frac{G}{c^3}\frac{I}{r^2}\frac{\mid\dot{f}_{\rm rot}\mid}{f_{\rm rot}}\right)^{1/2},
\end{equation}
\citep[see, e.g.,][]{2014ApJ...785..119A} where one is considering that the whole contribution to $\dot{f}_{\rm rot}$ is due to GW emission, i.e., the spin-down limit. This equation must be modified to take into account the magnetic braking \citep[see][]{2016arXiv160305975D,2016EPJC}. This can be done by writing  
\begin{equation}
\dot{\bar{f}}_{\rm rot} = \eta \dot{f}_{\rm rot}, 
\end{equation}
where $\dot{\bar{f}}_{\rm rot}$ can be interpreted as the part of $\dot{f}_{\rm rot}$ related to the GW emission brake. Consequently, the GW amplitude is given by
\begin{equation}
h^2 = \frac{5}{2}\frac{G}{c^3}\frac{I}{r^2}\frac{\mid\dot{\bar{f}}_{rot}\mid}{f_{rot}} =  \frac{5}{2}\frac{G}{c^3}\frac{I}{r^2}\frac{\mid\dot{f}_{\rm rot}\mid}{f_{\rm rot}} \, \eta . \label{heta}
\end{equation}
On the other hand, the GWs amplitude can also be written as follows
\begin{equation}
h = \frac{16\pi^2G}{c^4} \frac{I\epsilon f_{\rm rot}^2}{r}
\end{equation}
\citep[see, e.g.,][]{1983bhwd.book.....S}. By combining the two equations above one can obtain $\epsilon_{B}$ in terms of $P$, $\dot P$ (observable quantities), $\eta$ and $I$, namely
\begin{equation}
\epsilon = \sqrt{\frac{5}{512\pi^4} \frac{c^5}{G}\frac{\dot{P}P^3}{I}\eta}. \label{epet}
\end{equation}

Still concerning $\eta$, as discussed in detail by \cite{2016ApJ}, it can be also interpreted as the fraction of the rotation power ($\dot{E}_{rot}$) emitted in the form of GWs ($\dot{E}_{GW}$), or yet, the efficiency for GWs generation. Obviously, part of the rotation power is emitted in the form of electromagnetic radiation through magnetic dipole emission ($\dot{E}_{d}$).

Also, it is shown in \cite{2016ApJ} an useful equation relating $\eta$ to the pulsar dipole magnetic field, which is derived by recalling that the magnetic brake is related to $P$ ans $\dot{P}$, i.e. 
\begin{equation}
\bar{B}_0\sin^2\phi = \frac{3 I c^3}{4 \pi^2 R^6} P \dot{P}
\end{equation}
where $\bar{B}_0$ would be the magnetic field whether the breaking is purely magnetic. Since pulsars might also emit GWs, $B_0 < \bar{B}_0$ is a reasonable assumption. Thus, the equation relating $\eta$ to the pulsar dipole magnetic field reads \citep[see][for details]{2016ApJ}
\begin{equation}
\eta = 1 - \left(\frac{B_0}{\bar{B}_0} \right)^2.
\end{equation}
By substituting this last expression into equation \ref{eq:epsilonB} one obtains
\begin{equation}
\epsilon_B = \frac{3Ic^3}{4\pi^2GM^2R^2}P\dot{P} \left(1 - \eta\right) \kappa. \label{eek}
\end{equation}
In addition, by combining this equation with equation \ref{epet}, one has  
\begin{equation}
\eta = \frac{288}{5}\frac{I^3c}{GM^4R^4}\frac{\dot{P}}{P}\left( 1-\eta \right)^2 \kappa^2. \label{ek}
\end{equation}

Since $\epsilon$ and $\eta \ll 1$, one can readily obtain the following useful equations 
\begin{equation}
\epsilon_B \simeq \frac{3Ic^3}{4\pi^2GM^2R^2}P\dot{P}\kappa
\label{eek2}
\end{equation}
and 
\begin{equation}
\eta \simeq \frac{288}{5}\frac{I^3c}{GM^4R^4}\frac{\dot{P}}{P} \kappa^2. \label{ek2}
\end{equation}

Now, we are ready to calculate $\epsilon_{B}$, $\eta$ and the GW amplitudes for all pulsars with known $P$, $\dot{P}$, and estimated distances to the Earth, for given values of $M$, $R$, $I$ and $\kappa$. The next section is devoted to such an issue as well as the corresponding discussion of the results. 

\section{Calculations and discussions}
\label{sec:3}

\begin{figure}
\includegraphics[width=\linewidth,clip]{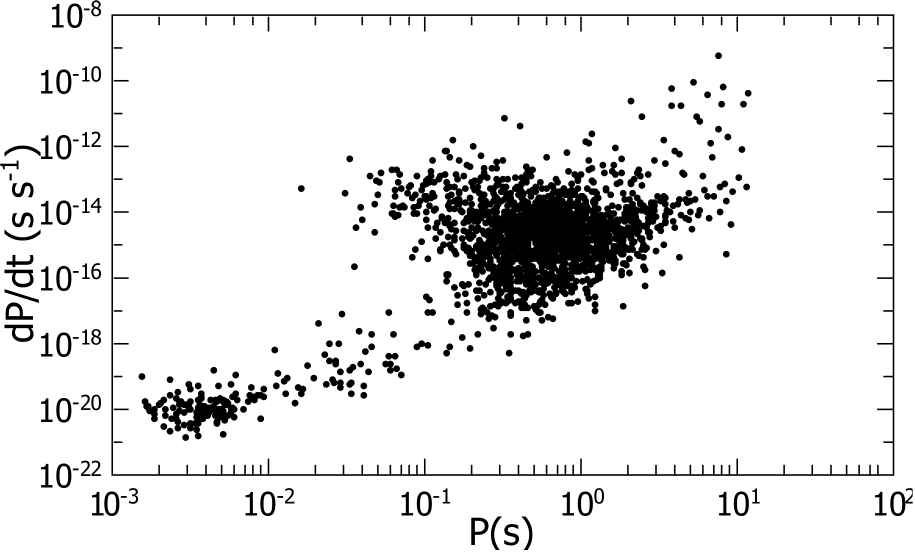} %fig1.pdf}
\caption{The $P$-$\dot{P}$ diagram for radio pulsars obtained from the ATNF Pulsar Catalog (http://www.atnf.csiro.au/people/pulsar/psrcat/)}\label{fig1}
\end{figure}
\begin{figure}
\includegraphics[width=\linewidth,clip]{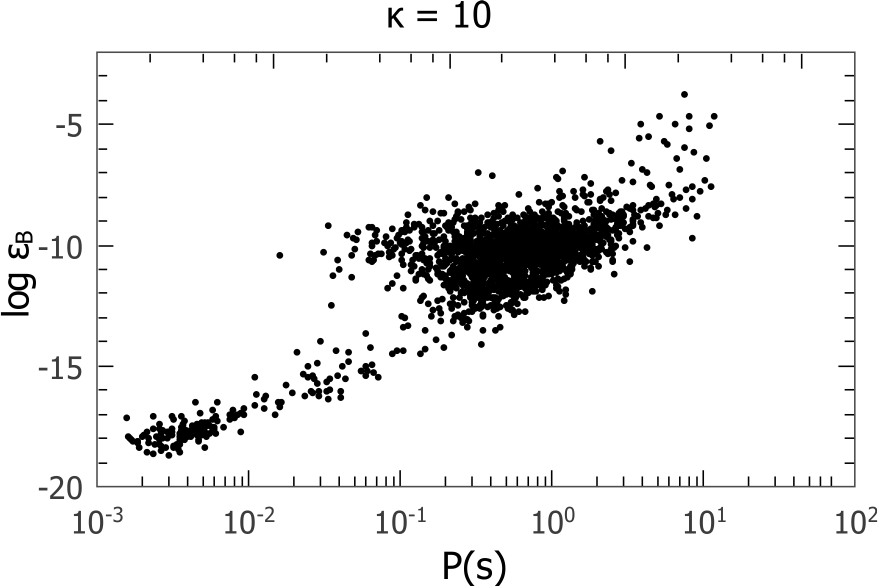} % fig2.pdf}
\caption{Correlation between the magnetic ellipticity and the period for all pulsars with known $P$, $\dot{P}$.}\label{fig2}
\end{figure}
\begin{figure}
\includegraphics[width=\linewidth,clip]{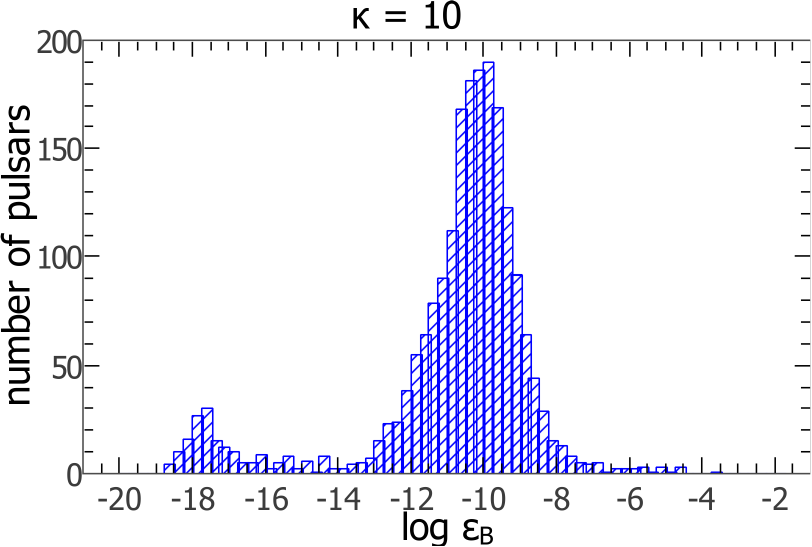}  % fig3.pdf} 
\caption{Ellipticity histogram for the 1964 pulsars of Table\ref{ta1} for $k=10$.}\label{fig3}
\end{figure}

The table with the necessary parameters used for the calculation of $\epsilon_{B}$, $\eta$ and $h$ as well as the adopted criteria for the selection of the pulsars can be found in the Appendix.
Fig.~\ref{fig1} is made with columns 2 and 3 of this very table, in which the pulsars' periods and their corresponding time derivatives (spin-down rate) are shown. Although being a well known diagram, it will be very useful for the discussions of the subsequent results. It is easily distinguishable two  pulsar populations: the millisecond pulsars in the lower left (with periods $P<10^{-2}\,\rm{s}$), composed by young or recently formed pulsars, and another class of pulsars with periods $10^{-1}<P<10^1\,\rm{s}$, composed probably by older pulsars which have already depleted a respectful fraction of their rotation power. 

On the last three columns of Table~\ref{ta1}, we present the results of our calculations, namely, $\epsilon_{B}$, $\eta$ and $h$. In order to perform these calculations, we also need to provide values for $M$, $R$, $I$ and $\kappa$. For the first three parameters, we are adopting fiducial values, namely, $M = 1.4 M_{\odot}$, $R = 10\, {\rm km}$, and $I = 10^{38} {\rm kg\,m^{2}}$. Regarding the distortion parameter $\kappa$, as already mentioned, it depends on the EoS and on the magnetic field configuration. In particular, we choose $\kappa = 10$, but values as high as $\kappa = 1000$ could be considered, although they are probably unrealistic~\citep[see e.g.,][for a brief discussion]{2006A&A...447....1R}. 

From equation \ref{eek2}, we find that $\epsilon_{B}$ is extremely small ($\sim10^{-19}-10^{-15}$) for the millisecond pulsars of Table~\ref{ta1}, even when an extremely optimistic case in which $\kappa \sim 1000$ is considered. Moreover, the ellipticity distribution assumes values of the order of $\sim 10^{-10}$ for the slowest pulsars. 
In fig.~\ref{fig2} we present $\log \epsilon_{B}$ versus $P$ for $\kappa = 10$ for all pulsars of Table~\ref{ta1}. The ellipticity for different values of $\kappa$, since $\epsilon_{B} \propto \kappa$, can be readily obtained. 

An interesting histogram can also be made from Table~\ref{ta1}, namely, the number of pulsars for $\log \epsilon_{B}$ bin (see fig.~\ref{fig3}). It is worth noticing the high number of pulsars concentrated around $\sim 10^{-10}\, (10^{-8})$  for $ k = 10 \,(1000)$. 

\begin{figure}
\includegraphics[width=\linewidth,clip]{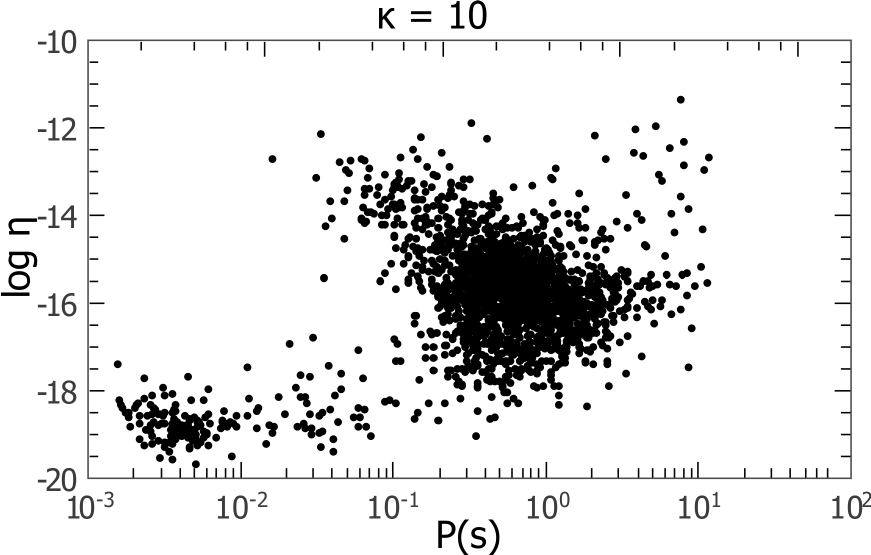} % fig4.pdf}
\caption{Correlation between the efficiency and the period for all pulsars with known $P$, $\dot{P}$.}\label{fig4}
\end{figure}

\begin{figure}
\includegraphics[width=\linewidth,clip]{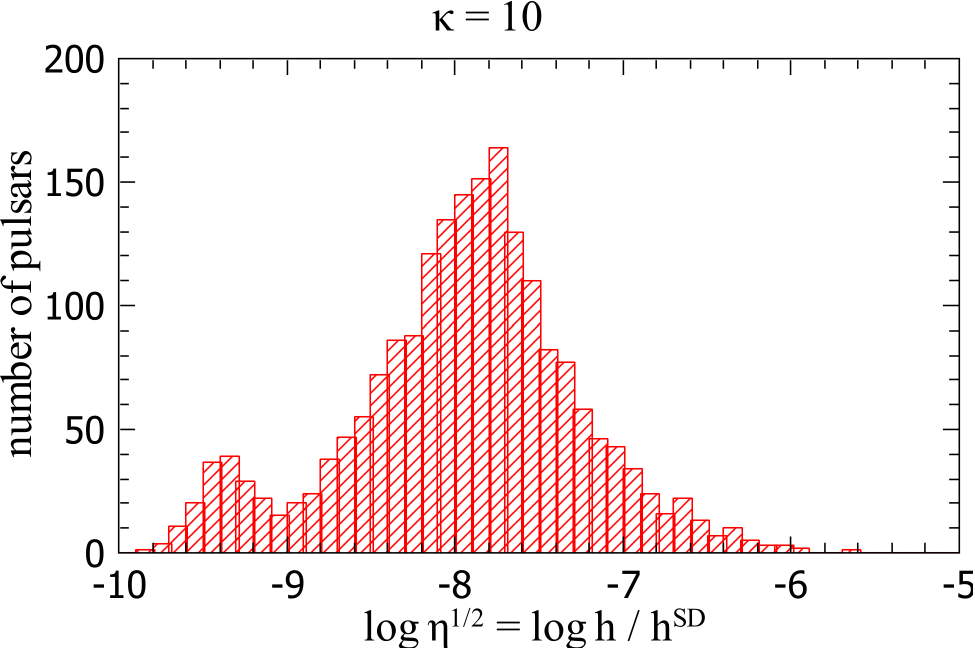} % fig5.pdf}
\caption{Histogram of the efficiencies (or the spin-down ratio) for the 1964 pulsars of Table\ref{ta1} for $k=10$.}\label{fig5}
\end{figure}

A similar analysis can be made for $\eta$ by means of Eq.~\ref{ek2}. In fig.~\ref{fig4} we present $\log \eta$ versus $P$ for $\kappa = 10$ for all pulsars of Table~\ref{ta1}. Notice that $\eta$ is also extremely small, even if one consider $\kappa = 1000$. Also, an histogram for the number of pulsars for $\log \eta^{1/2}$ can be seen in fig.~\ref{fig5}. This quantity ($\eta^{1/2}$) is equivalent to  the ratio $h/h^{\rm{SD}}$, i.e., the spin-down ratio. One may notice a peak in the $\eta$ histogram at $10^{-16} - 10^{-15}$ for the pulsars of Table \ref{ta1}. As for $\epsilon_{B}$, $\eta$ for different values of $\kappa$ can be easily obtained, since $\eta \propto \kappa^2$. Thus, even for $\kappa = 1000$ the peak in the histogram would be around $10^{-12} - 10^{-11}$.
Before proceeding, it is worth stressing that the two pulsar populations aforementioned also clearly appear in figs~\ref{fig2} - \ref{fig5}.

\begin{figure}
\includegraphics[width=\linewidth,clip]{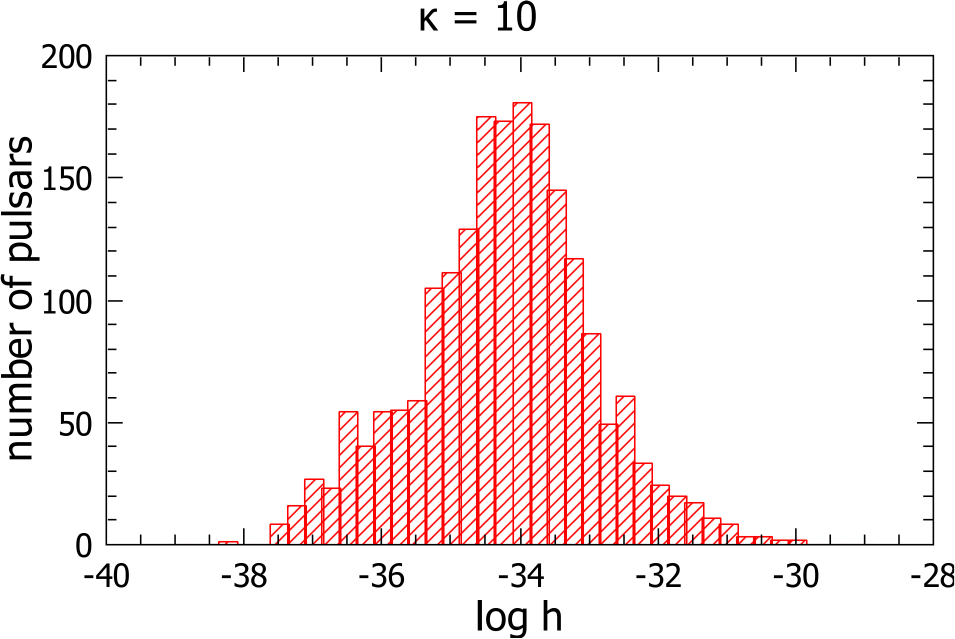} % fig6.pdf}
\caption{Histogram of the amplitude of GWs for all pulsars  with known $P$, $\dot{P}$.}\label{fig6}
\end{figure}

These extremely small values of $\epsilon_{B}$ and $\eta$ have very important consequences as regards the detectability of GWs generated by the pulsars, whether the ellipticity is mainly due to the magnetic dipole of the pulsars themselves. 
Our calculations show that the GW amplitudes for most of these pulsars are at best seven orders of magnitude smaller than those obtained by assuming the spindown limit (SD), see Fig.~\ref{fig6}. 
Notice that, even considering an extremely optimistic case, the value of the ellipticity is at best $\epsilon_{B} \sim 10^{-5}$ (for PSR J1846-0258) and the corresponding efficiency $\eta \sim 10^{-8}$. Thus, the GW amplitude even in this case would be four orders of magnitude lower than the amplitude obtained by assuming the spindown limit ($\eta =1$).

Still regarding  fig.~\ref{fig5}, a recent result from the searches for GWs from 200 pulsars by using the aLIGO first observation run data \citep[see figure 4 of][]{arXiv:1701.07709} presents a similar distribution for $h/h^{\rm{SD}}$, but displaced by orders of magnitude. Although, in such a work, much higher ellipticities are obtained and no discussion is done about the mechanism responsible for that. Yet those values for the ellipticities should be considered as upper limits since no continuous GW has been detected. In our present work, as already mentioned, we have considered that the ellipticity is caused exclusively by magnetic dipole and showed that such a mechanism alone could not generate a much strong star deformation. In our view, other mechanism or mechanisms should be associated to the magnetic field to make possible those putative high ellipticities to be present, e.g., high precession and/or non conventional EoS.

For a matter of comparison, some authors argue that a fiducial
upper limit for the ellipticity would be around $\epsilon \sim 10^{-6}$, considering asymmetries supported by anisotropic stress built up during the crystallization period of the crust \citep[see, e.g.,][and references therein]{2008PhLB..668....1K}.

Finally, since the predicted GW amplitudes are extremely small for all pulsars of Table \ref{ta1}, and thousands of years of observing time would be needed, even advanced detectors such as aLIGO and AdV, and the planned ET would not be able to detect these pulsars, whether the ellipticity is of magnetic dipole origin.

%\bigskip

%%%%%%%%%%%%%%%%%%%%%%%%%%%%%%%%%%%%%%%%%%%%%%%%%%%%%
% \begin{figure}
% \includegraphics[width=\linewidth,clip]{EpsilonEta.pdf}
% \caption{...}\label{fig1}
% \end{figure}

% \begin{figure}
% \includegraphics[width=\linewidth,clip]{dotPepsilon.pdf}
% \caption{...}\label{fig5}
% \end{figure}

% \begin{figure}
% \includegraphics[width=\linewidth,clip]{dotPeta.pdf}
% \caption{...}\label{fig6}
% \end{figure}
%%%%%%%%%%%%%%%%%%%%%%%%%%%%%%%%%%%%%%%%%%%%%%%%%%%%%%%

\section{Conclusions and final remarks}
\label{sec:4}

In this paper, we extend our previous studies ~\citep{2016ApJ} in which we considered the role of magnetic ellipticity on the braking index and on the pulsar distortion. It is well known that for strong magnetic fields ($\sim10^{12}-10^{15}$~G), the equilibrium configuration of a neutron star can be distorted by magnetic tension.
Such magnetic fields could not generate enough deformation to lead to high values of ellipticity. The presence of a magnetic field exceeding the magnetar strength ($\sim10^{16}$ G) could account for an ellipticity of $\sim10^{-4}$. Recall that much higher values of magnetic fields could violate the Virial Theorem and disrupt the star equilibrium~\citep[see, e.g.,][]{2009MNRAS.395.2162L}.

Here we consider the role played by the magnetic dipole field on the deformation of the known pulsars and its consequences as regards the generation of GWs. In particular, we obtained useful equations, \ref{eek2} and \ref{ek2}, with which one can calculate $\epsilon_{B}$ and $\eta$ in terms of $I$, $M$, $R$, $\kappa$ and the observable quantities $P$ and $\dot P$, and distances to the Earth. Moreover, the amplitudes of GWs can be readily calculated. It is worth stressing that this is the first time in literature that such a study is done for a large sample of pulsars.

Regarding the GWs generated by the pulsars, our calculations show that the amplitudes are extremely small, as result the prospect for their detection even for aLIGO, AdV, and the planned ET, would be pessimistic. This conclusion is obviously dependent on the mechanism that generates the ellipticity. Whether there is some other mechanism that could generate substantially larger ellipticities the prospects for the detection of GWs emitted by the pulsars of Table \ref{ta1} could be much less pessimistic. 

\section*{Acknowledgments}
J.C.N.A thanks FAPESP (2013/26258-4) and CNPq (308983/2013-0) for partial support. J.G.C. acknowledges the support of FAPESP (2013/15088-0 and 2013/26258-4).  C.A.C. acknowledges PNPD-CAPES for financial support.

%\bibliography{ref}   % name your BibTeX data base
%%%%%%%%%%%%%%%%%%%%%%%%%%%%%%%%%%%%%%%%%%%%%%%%%%



%\bibliography{ref}
%\bibliographystyle{apj}

\newpage

\onecolumn %to fix longtable problem
\appendix 
\label{app}

Table \ref{ta1} presents the periods ($P$), their first derivatives ($\dot{P}$), and the estimated distances ($d$) from Earth of 1964 pulsars drawn from the ATNF Pulsar Catalog (http://www.atnf.csiro.au/people/pulsar/psrcat/). Another selection criteria is that the pulsars must have only an unique measured value for $\dot{P}$. It is also presented $\eta$, $\epsilon$ and $h$ values calculated from the appropriate equations which can be found in Section \ref{sec:3}.  

\begin{longtable}{lcccccc}
\caption{$P$, $\dot P$ and $d$ for pulsars drawn from the ATNF catalog, with an unique measured value for $\dot P$, and the correspondent $\epsilon$, $\eta$ and $h$ for $\kappa = 10$.}
\label{ta1} \\
\hline
Pulsar & $P$~(s) & $\log \dot{P}$ & d (kpc) & $\log \epsilon$ & $\log \eta$ & $\log h$ \\ 
\hline
\endfirsthead
\multicolumn{7}{c}%
{{\bfseries \tablename\ \thetable{} -- continued from previous page}} \\
\hline 
Pulsar & $P$~(s) & $\log \dot{P}$ & d (kpc) & $\log \epsilon$ & $\log \eta$ & $\log h$ \\
\hline
\endhead
\hline 
\multicolumn{7}{r}{{Continued on next page}} \\ 
%\hline
\endfoot
\hline
\endlastfoot
J0006+1834 & 0.693748 & -14.6778 & 0.7 & -10.1931 & -15.7481 & -33.0943 \\ 
J0007+7303 & 0.315873 & -12.4437 & 1.4 & -8.30073 & -13.1724 & -30.8195 \\ 
J0014+4746 & 1.2407 & -15.2487 & 1.82 & -10.5116 & -16.5715 & -34.3327 \\ 
J0026+6320 & 0.318358 & -15.8239 & 13.6 & -11.6775 & -16.556 & -35.1906 \\ 
J0030+0451 & 0.004865 & -19.9914 & 0.3 & -17.6609 & -18.9076 & -35.8858 \\ 
J0034-0534 & 0.001877 & -20.3036 & 0.98 & -18.3867 & -18.8063 & -36.2985 \\ 
J0034-0721 & 0.942951 & -15.3893 & 1.03 & -10.7714 & -16.593 & -34.1069 \\ 
J0040+5716 & 1.11823 & -14.5406 & 4.48 & -9.84862 & -15.8183 & -33.9706 \\
\end{longtable}
%\end{table}

\end{document}